# A Narrative Literature Review and E-Commerce Website Research

K.M. Rahman[1]

[1]View Soft Inc., Ashburn, Virginia, USA

## Abstract

In this study, a narrative literature review regarding culture and e-commerce website design has been introduced. Cultural aspect and e-commerce website design will play a significant role for successful global e-commerce sites in the future. Future success of businesses will rely on e-commerce. To compete in the global e-commerce marketplace, local businesses need to focus on designing culturally friendly e-commerce websites. To the best of my knowledge, there has been insignificant research conducted on correlations between culture and e-commerce website design. The research shows that there are correlations between e-commerce, culture, and website design. The result of the study indicates that cultural aspects influence e-commerce website design. This study aims to deliver a reference source for information systems and information technology researchers interested in culture and e-commerce website design, and will show less-focused research areas in addition to future directions.

## 1. Introduction

Electronic commerce (e-commerce) is a fairly new idea, and it is very common practice nowadays for businesses to conduct trade over the Internet. There are various advantages to e-commerce (e.g., lower cost, convenience). E-commerce can simply be defined as buying and selling merchandise or services online. Most successful businesses today have their own websites. Today, it is possible to conduct business nationally and globally with a click of a fingertip due to the worldwide use of the Internet. To be successful in the global marketplace, businesses need to develop culturally friendly e-commerce websites. When conducting business online, factors such as region and culture, web content accessibility, ease of use, secure authentication, payment, fraud detection, performance, trust, stability, technology, and convenience are vital to the businesses' and consumers' satisfaction and interest. This is a less focused research area and needs significant attention. E-commerce, therefore, is an important area for research and needs further investigation.

This study focuses on the correlation between e-commerce, culture, and website design, and the article begins with a literature review mostly related to these areas. At first, the article gives a brief definition of e-commerce. Next, the article introduces the literature review, including 4 types of literature review methods, and research methodology for this paper, followed by the resulting outcomes, findings of new significance, and limitations. Finally, the paper is summed up in a conclusion.

K. M. RahmanK. M. Rahman

## 2. Research Questions

E-commerce already has a huge impact globally, and it has been an important instrument for businesses to use for reaching out to the global consumer community. The study aims to conduct a narrative literature review and then—based on that literature review—find out the answers to the following questions:

- Are there any correlations between e-commerce, culture, and website design?
- Does culture influence e-commerce website design?

## 3. E-commerce Website Research Literature Review

Internet has major influence on the globe because it can serve billions of users all over the world. Thousands of local and global networks including private, public, academic, business, and government networks, all contribute to the creation of the Internet (Yongrui et al. 2014). Internet has opened the door for e-commerce. This section provides a brief overview of e-commerce, including what it is and how it differs from existing related ideas.

The use of e-commerce is growing as a way to conduct business (Ngai and Wat, 2002). The initial phase of conducting research regarding e-commerce is to explain the e-commerce idea. Table 1 describes e-commerce from diverse viewpoints.

| Table 1: Definition of E-commerce | |
|---|---|
| Definition | Reference |
| E-commerce simply can be defined as doing business online. | (Khurana and Mehra, 2015) |
| E-commerce stands for electronic commerce. It is trading in goods and services using computer networks, such as the Internet. | (Shahriari et al., 2015) |
| E-commerce denotes the paperless exchange of trade information using electronic data interchange, electronic mail, electronic bulletin boards, electronic funds transfer, the Web, and other network-based technologies. | (Bhalekar et al., 2014) |
| E-commerce indicates the buying and selling of goods and services through the Internet. | (Lim, 2014) |
| E-commerce is the interaction among communication systems, data management systems, and security, which together exchange commercial information in relation to the sale of goods or services. | (Nanehkaran, 2013) |
| E-commerce can be defined as buying and selling merchandise or services on the Internet or other networks. | (Khoshnampour and Nosrati, 2011) |
| E-commerce is the manner of commercial trades over electronic networks. | (Organization for Economic Co-operation and Development (OECD), 2002) |
| Electronic commerce is the sharing of business information, sustaining business relationships, and doing business dealings by means of telecommunications networks. | (Zwass, 1996) |

*Prepared by Khan Mashiur Rahman*

E-commerce, therefore, is the purchase and sale of goods and services over the Internet. Businesses have their websites which allow consumers to browse products and services. Consumers then purchase the goods and services using various methods (e.g., credit card, debit card, electronic check, PayPal) via the website. Finally, the businesses ship the orders to the consumers' doorsteps. Consumers are also able to track the status of the product or service ordered as needed. All of these activities are conducted online.

There are many types of e-commerce, but the most common are as follows:

- B2B (Business-to-Business): B2B involves trade dealings between businesses where sale of merchandise is made to an interim purchaser (Pandey and Agarwal, 2014). B2B e-commerce deals with interactions between and among businesses. About 80% of e-commerce is in this category (Gupta, 2014). One example of B2B is a business transaction between a manufacturer and a wholesaler, or between a wholesaler and a retailer.

- B2C (Business-to-Consumer): In this model the trade is conducted directly between company and consumers via a website (Dan, 2014). In the B2C model, businesses or organizations trade goods or services to customers over the Internet for consumers' own use.

- C2C (Consumer-to-Consumer): This is the fastest-growing type of e-commerce. Businesses simply provide a platform (i.e., www.ebay.com) to advertise the product, and buyers can buy the product directly from the seller (Pandey and Agarwal, 2014). In the C2C model, businesses facilitate the setting where consumers buy and sell goods and services directly to each other.

- C2B (Consumer-to-Business): In a C2B model, customers sell goods and services to businesses, and the corporations purchase the products and services (Nemat, 2011). Consumers provide value, and the businesses consume that value.

Internet has major influence on the globe since it can serve billions of users all over the world. Thousands of local and global networks including private, public, academic, business, and government networks, all contribute to the creation of the Internet.

E-commerce is overwhelmingly affecting both national and global economy. Recent research studies indicate that e-commerce has a positive influence on the development of countries' overall economies, and this contribution to economies will continue to grow (Shahriari et al., 2015). There is no doubt that the overall sales and productivity of an organization will increase due to the trend of this technological enhancement.

The idea of culture is very complex. E-commerce websites are always changing. The global Internet usage is also growing considerably. Most of the e-commerce websites are in English, but the majority of the online users are non-English speaking; other languages such as Chinese, Spanish, Japanese, and French currently have substantial Internet usage (Singh et al., 2010). Most information on the website is presented by elements such as colour, images, logos, shapes, frame, links, banner, buttons, animation, splash windows, and white spaces (Asimionoaei, 2009; Punhani and Batra, 2014). It is important that e-commerce website designers consider the local culture when designing the website to reflect the local users' emotions, thoughts, desires, and most importantly, the culture. The study indicates that different countries' e-commerce website users have diverse preferences and motives to use global e-commerce websites (Punhani and Batra, 2014). Writing style, language option, navigation models, history, ethics, security and trust, gender, public policy, advertisement, awareness, human interaction, procedural compatibility, corporate structure, and prior e-commerce experience can also influence the e-commerce website user's view (Gould, 2006; Akhter, 2016; Brdesee et al., 2012; Dehkordi et al., 2011; Shahdadnejad and Nakhaie, 2011; Shahriari et al., 2015). All of these features and cultural aspects are very important factors for successful e-commerce websites.

The use of these website features to design an e-commerce website may be suitable to some local cultural groups but may not be appropriate by other global viewers. For the website design characteristics, the research includes colour usage, page layout, site content, and the nature of website interactions. These characteristics were used to determine whether there was any indication of cultural influence among the selected website design. E-commerce website designers need to think through the influence of local culture and how the local culture motivates users' perceptions. For example, there is a greater use of red colour in Chinese websites (Lo and Gong, 2005). The language used in an e-commerce website is another important factor for a successful business. E-commerce sites are more successful when they offer sales in the customer's language. The study indicates that if the e-commerce website is presented in the consumer's language, the number of individuals visiting the e-commerce website will be four times higher than the usual number of visitors (Shahdadnejad and Nakhaie, 2011). Another consideration in the study was the type of advertisements placed on the e-commerce website. Advertisements appropriate in one culture may not be suitable for another culture (Shahdadnejad and Nakhaie, 2011). The study also indicates that customers have concerns regarding security and privacy of personal data when conducting business online (Li et al., 2011; Wang, et al., 2009; Wang, et al., 2002; Wang, et al., 2001; Wang, et al., 2005). To overcome some of the security issues, electronic cash has been playing a significant role in e- commerce (Wang and Zhang, 2011). The research indicates that all of these issues significantly influence users' desire to conduct business through e-commerce websites.

Most of the research studies done on this topic are narrowly focused on culture and how the cultural factors are influencing e-commerce. The adoption of e-commerce across cultures is very important for countries' economic development (Hofstede, 1980; Hofstede, 1991; Flood, 2014; Kang, 2005; Belkhamza and Wafa, 2014). It is obvious that to compete in the global market, the e-commerce website designers need to consider the influences of the regional and local cultures when designing the e-commerce website (Lin, 2015). This will allow the e-commerce website to reach out to the global consumer community to maximize the revenue for its business.

Australia, the United States, Canada, and the United Kingdom will be some ideal developed countries in which to conduct the research. All of these countries have state-of-the-art technology infrastructures. People of these developed nations are very comfortable with technology and willing to conduct businesses through e-commerce websites (Gould, 2006). Therefore, finding data to analyse for research will be effective.

Bangladesh is an ideal developing country in which to conduct part of the research. The technology infrastructure of Bangladesh is not as advanced as that of the developed countries; therefore, it will be a great challenge to conduct research in Bangladesh, but it will also be interesting. The e-commerce industry is still new in Bangladesh. There is huge potential for Bangladesh to compete in the global marketplace by means of e-commerce. Unfortunately, Bangladesh is lagging way behind its competition in the global e-commerce market, especially in the B2C dimension (Mohiuddin, 2014). Competing in the global e-commerce marketplace is, therefore, going to be another challenging aspect for Bangladesh. As a developing country, Bangladesh has many issues (e.g., high unemployment, over population, less resources). E-commerce in Bangladesh will certainly benefit some of these areas.

A summary of research studies conducted regarding correlations among e-commerce, culture, and website design are listed in Table 2, which also presents studies on how cultural factors are influencing e-commerce website design.

| Table 2: A Summary of Research Regarding Correlations Between E-commerce, Culture, and Website Design | |
|---|---|
| Author(s) and Year | Research Outcome |
| (Punhani and Batra, 2014) | The comparative analysis shows that it is important that e-commerce website designers consider the local cultural aspects when designing the website to reflect the local users' emotions, thoughts, desires, and most importantly the culture. The important conclusion from the case study was that different countries' e-commerce website users have diverse preferences and motives to use global e-commerce websites. |
| (Shahdadnejad and Nakhaie, 2011) | The findings of their case study indicated that if the e-commerce website is presented in the consumer's language, the number of individuals visiting the e-commerce website will be four times higher than the usual number of visitors. Another point in the study was the type of advertisements placed in the e-commerce website, which are also an important aspect to consider. The advertisements appropriate in one culture may not be suitable for another culture. |
| (Asimionoaei, 2009) | The research shows that e-commerce web page design and colours used on the web page have diverse emotional and social consequences in the global cultural community. The language is another important factor along with culture when designing an e-commerce website. E-commerce businesses need to develop their websites with choices for numerous languages. This study concluded that to have a successful e-commerce business, an e-commerce website development process must pay sincere attention to global culture. |
| (Lo and Gong, 2005) | The research found that there is a need for e-commerce website designers to think through the influence of local culture and how the local culture motivates a user's perception. For example, there is a greater occurrence of red colour in Chinese websites. |

*Prepared by Khan Mashiur Rahman*

A summary of selected studies in culture, e-commerce website design, and narrative review is presented in Table 3, which gives a brief overview of all the materials reviewed and analysed in this study. Table 3 presents authors' names and the years the articles were published, topics studied, and important factors examined.

| Table 3: A Summary of Selected Studies in Culture, E-commerce Website Design, and Narrative Review | | |
|---|---|---|
| Author(s) and Year | Explanation of Study | Important Feature Examined in the Study |
| (Akhter, 2016) | Cultural dimensions of behaviours towards e-commerce | The effect of e-commerce in Saudi Arabian culture |
| (Khurana and Mehra, 2015) | E-commerce opportunities and challenges | The concept of e-commerce, e-commerce scenario in India, and opportunities and various challenges in e-commerce |
| (Lin, 2015) | Facilitating cultural and creative industries to engage the Internet era | Cultural and Creative Industries (CCI) and e-commerce strategies in the perspective of international trade setting |
| (Shahriari et al., 2015) | Uncertainty avoidance role on e-commerce acceptance across cultures | How Malaysia and Algeria are adopting e-commerce |
| (Bhalekar et al., 2014) | Study of e-commerce | Benefits and limitations of e-commerce |
| (Belkhamza and Wafa, 2014) | E-commerce impact on global market | Global influence of e-commerce and its benefits |

A Narrative Literature Review and E-Commerce Website Research

**Table 3: A Summary of Selected Studies in Culture, E-commerce Website Design, and Narrative Review**

| Author(s) and Year | Explanation of Study | Important Feature Examined in the Study |
|---|---|---|
| (Dan, 2014) | Electronic commerce | Discussed the historical and current activities in e-commerce process |
| (Flood, 2014) | Humanism and culture | Investigated several ways social scientists tried to define and categorized various cultures |
| (Gupta, 2014) | The role of e-commerce in today's business | Role of e-commerce in Today's business |
| (Lim, 2014) | Adoption of e-commerce | Factors that affect the acceptance of e-commerce |
| (Mohiuddin, 2014) | Overview of the e-commerce in Bangladesh | Overall e-commerce websites and business to consumer category of Bangladesh |
| (Pandey and Agarwal, 2014) | E-commerce transactions: An empirical study | E-commerce trades, its various types and classes, its effectiveness, and the key features |
| (Punhani and Batra, 2014) | Understanding cultural variations of e-commerce websites in a global framework | Common features to compare website design across different country |
| (Yongrui et al. 2014) | Data centric view of Internet of Things | Internet of Things from data centric perspectives |
| (Nanehkaran, 2013) | Introduction to electronic commerce | Principles, definitions, history, frameworks, steps, models, advantages, barriers, and limitations of e-commerce |
| (Brdesee et al., 2012) | Cultural means obstructing e-commerce adoption in a Saudi industry | Saudi travel businesses and various issues affecting the acceptance of e-commerce |
| (Dehkordi et al., 2011) | Factors that influence the acceptance of e-commerce in developing countries | Whether there are correlations between national culture, gender, and prior e-commerce experience on certain factors |
| (Khoshnampour and Nosrati, 2011) | Overview of electronic commerce | Various applications of e-commerce |
| (Li et al., 2011) | Privacy-aware access control with trust management in web service | Privacy is a major concern in mobile commerce |
| (Nemat, 2011) | Taking a look at different types of e-commerce | Various types of e-commerce |
| (Shahdadnejad and Nakhaie, 2011) | Culture's role in development of e-commerce | Technological improvement and use of e-commerce in our everyday lives |
| (Wang and Zhang, 2011) | Untraceable off-line electronic cash flow in e-commerce | Electronic cash and e-commerce security |
| (Singh et al., 2010) | Global e-commerce importance | The significance of e-commerce in the global economy perspective |
| (Asimionoaei, 2009) | Cultural issues in website design | Consideration of culture when designing an e-commerce website |
| (Wang et al., 2009) | Effective collaboration with information sharing in virtual universities | Information sharing in secure manner for virtual university collaborative environments |
| (Gould, 2006) | Study on cultural values | Examination of eight websites to find possible rules for cross |

**Table 3: A Summary of Selected Studies in Culture, E-commerce Website Design, and Narrative Review**

| Author(s) and Year | Explanation of Study | Important Feature Examined in the Study |
|---|---|---|
| | and e-commerce design | cultural website design |
| (Kang, 2005) | Considering the cultural issues of e-commerce website design | Investigated cultural differences in web design and designers' views |
| (King and He, 2005) | The role and method of meta-analysis in literature review | Four ways a literature review can be conducted |
| (Lo and Gong, 2005) | Cultural impact on the design of e-commerce websites | The level of cultural influence on the website design |
| (Wang et al., 2005) | A flexible payment scheme and its role-based access Control | Security and privacy concern when purchasing online |
| (Organization for Economic Co-operation and Development (OECD), 2002) | Evaluating the information economy | The role of information and communication technologies (ICT) in economic development |
| (Ngai and Wat, 2002) | Literature review and classification of electronic commerce examination | The study conducted on four main categories: application areas, technological issues, support and implementation, and others |
| (Wang et al., 2002) | Ticket-based service access scheme for mobile users | In the use of mobile devices, security is one of the main concerns |
| (Webster and Watson, 2002) | Writing a literature review | Investigated how an effective literature analysis is a vital component for academic research project |
| (Wang et al., 2001) | Customer scalable | Security and privacy of personal data when |
| | security | conducting business online |
| (Zwass, 1996) | Electronic commerce: structures and issues | Discussed a categorized structure of e-commerce expansion and analysis, ranging from the widespread area telecommunications infrastructure to electronic marketplaces and electronic hierarchies enabled by e-commerce |
| (Hofstede, 1991) | Cultures and organizations | Established that national cultures vary in five dimensions |
| (Guzzo et al., 1987) | Analysing Meta-analysis | The study examined meta-analysis, a quantitative form of literature review |
| (Hofstede, 1980) | Consequence of culture | How national culture impacts behavioural intent |

*Prepared by Khan Mashiur Rahman*

A careful consideration of Table 3 establishes that e-commerce is an important area for research and needs continued investigation. There are challenges in adapting e-commerce for various countries. Also, the national culture will play a major role in conducting successful e-commerce business in the global e-commerce marketplace. It is obvious that e-commerce has made and will continue to make a significant impact in the global economy.

## 4. E-commerce Website Research Literature Review

An appropriate literature review is a vital part of every academic research assignment. A related and effective literature review lays a strong groundwork for progressing knowledge (Webster and Watson, 2002). The study shows that deficiency of review articles has been obstructing the development of information systems arena (Webster and Watson, 2002).

A literature review can be conducted in many ways; Figure 1 illustrates four methods of literature review: Narrative Review, Descriptive Review, Vote Counting, and Meta-Analysis (King and He, 2005). These four methods are positioned on a qualitative-quantitative scale to demonstrate their dissimilar concentrations).

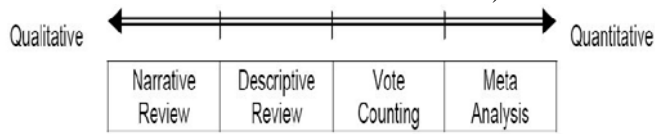

**Figure 1.** Literature Review Methods on a Qualitative–Quantitative Scale (King and He, 2005)

In a *narrative review*, the review is conducted in an old-fashioned manner. There are no uniform processes. The review is done by verbally describing the earlier studies, concentrating on theories and frameworks, elementary factors, and their research outcomes, with regard to a hypothesized correlation (King and He, 2005). Narrative review depends mostly on the reviewer's individual preference. It is not unusual for two reviews to reach rather different conclusions from the same literature (Guzzo et al., 1987).

A *descriptive review* introduces some quantification, often in a form of frequency analysis (e.g., publication time, research methodology, research outcomes). Descriptive review concentrates on identifying an interpretable pattern from the existing literature (Guzzo et al., 1987). Initially, a reviewer needs to perform a thorough literature search to gather adequate related papers in the research area. The reviewer then gives a single study as one data record and recognizes trends and patterns among the papers surveyed (King and He, 2005*)*.

*Vote counting* usually explains the main relationships by combining individual research results (King and He, 2005). In vote counting, calculation is made of the occurrence with which existing research discoveries support a specific proposal. A sequence of tests is conducted to create perceptions.

*Meta-analysis* is a statistical blend technique that provides support for a research topic by uniting and examining the quantitative outcomes of many experiential studies (King and He, 2005). In meta-analysis, qualitative studies have to be omitted because of the technique's vastly quantitative nature (King and He, 2005).

A narrative literature review method was most applicable for the existing phase of this research. The narrative literature review was suitable in this study because there was no uniform process followed by the author when conducting the literature reviews for each of these documents and materials. The narrative literature review method allowed the author to express author's own personal view. The author analysed and summarized the earlier studies conducted. The process for conducting this narrative review is defined in the next section.

## 5. Research Methodology

In this research, the goals are to reveal a background of e-commerce, culture, website design, and narrative review as a developing research area and to demonstrate a vision to guide future growth. This article is based on documents and materials associated with e-commerce, culture, website design, and narrative review. The documents and materials were collected from journals, research papers, and published books.

*Literature Exploration Scopes*

To conduct literature review, it is first necessary to identify related literature through computer and manual searches (Ngai and Wat, 2002). In this research, related literature from various journals and conferences was identified by using Google's search engine. The Google search was conducted by using a keyword search with the phrases 'e-commerce,' 'culture,' 'website,' and 'narrative review.' E-commerce is a new concept, so it is more viable to get current information from the Internet rather than from the library when conducting research on this topic.

*Literature Filtering Method*

At first, a total of 67 literature sources were identified and scanned for relevant topics using the Google search engine. The 67 articles included diverse issues such as e-commerce, developed and developing countries, culture, website design, localization, globalization, literature review methods, adaption of e-commerce, trust, security, privacy, policy, gender, international trade, information systems, information technology and distributed systems. These 67 journal articles provided an overall background regarding e-commerce. The second step was to scan abstracts and read complete manuscripts if needed for each article. Finally, 39 articles were studied thoroughly for the purposes of this study. Out of 67 articles, a total of 28 articles were discarded due to various reasons (e.g., duplicate or irrelevant information).

## 6. New Significant Findings and Research Synthesis

The study shows that overall use of Internet and e-commerce is growing considerably throughout the world. English is the language used in most of these e-commerce websites; the majority of these Internet users, however, are non-English speaking, and other languages (e.g., Chinese, Spanish, Japanese, and French) also have considerable Internet presence. E-commerce sites are successful when they offer to sell in the customer's language. The findings of the case study indicate that if the e-commerce website is presented in consumer's language, the number of individuals visiting the e-commerce website will be four times higher than the usual number.

It is, therefore, very important that e-commerce websites be culturally friendly (e.g., greater use of red colour, active banners, and animations in Chinese websites; fewer colours, active banners and animations in American websites). The types of advertisement placed on the e-commerce websites are also important aspects to consider, because advertisements appropriate in one culture may not be suitable for another culture.

The future success of businesses lies in e-commerce. With the use of e-commerce, consumers now have the ability to shop online anytime and anywhere in the globe. The invention of computer and Internet led the people to spend less time on other forms of entertainment such as television, radio, and newspapers. Consumers are spending more time online, and the majority of that time is spent on social media websites such as Facebook and Twitter. These computer users are potential e-commerce customers for businesses globally. To compete and to be successful in the global e-commerce marketplace, therefore, organizations need to focus on designing their e-commerce websites in ways that are conveniently accessible throughout the world. As a result of this development, organizations also need to consider localization and cultural aspects when designing their e-commerce websites. From the developing countries' perspective, infrastructure, cost, security, trust, and political issues are some of the major barriers for the overall growth of e-commerce.

This study indicates that there are correlations between e-commerce, culture, and website design. The local cultures influence consumer behaviour as to how the e-commerce websites are utilized and browsed. Culture expressly impacts the users' desire to conduct business through e-commerce websites. The research also indicates that cultures do influence e-commerce website design.

Finally, the current study confirmed and reinforced the research questions. There are correlations between e-commerce, culture, and website design, which are confirmed in Table 2 and Table 3. The research shows that it is important that e-commerce website designers consider the local cultural aspects when designing the website to reflect the local users' emotions, thoughts, desires, and most importantly the culture. The different countries' e-commerce website users have diverse likings and reasons to use global e-commerce websites. E-commerce companies need to develop its websites with choices for several languages.

Furthermore, the second research question was answered in that cultural aspects influence the ecommerce website design; this can be shown in Table 2 as well. The research shows that if the e-commerce website is presented in the consumer's language, the number of users visiting the e-commerce website will be four times higher than the usual number of visitors. Another important aspect in the research was the type of advertisements placed in the e-commerce website, which are also an important aspect to consider. The advertisements appropriate in one culture may not be suitable for another culture. The research also shows that e-commerce web page design and colours used on the web page have diverse emotional and social consequences in the local culture. For example, there is a greater occurrence of red colour in Chinese websites.

## 7. Conclusion and Future Work

This study has a few limitations. First, the sample size was small and mostly based on academic and refereed journal articles. The sample, therefore, might not reflect the complete view and solid evidence regarding correlation among e-commerce, culture, and website design. Second, no quantitative analysis was conducted to back up the study. Third, the search criteria might be imperfect, because some papers may not have the terms 'e-commerce,' 'culture,' 'website design,' or 'narrative review' in the abstract or keyword list, or the keyword list may be missing. Finally, a narrative review method was used, so the review is mostly dependent on the reviewer's individual preferences. It is likely that two reviewers could reach different conclusions from the same literature.

We live in the technology era. Every country in the globe is investing in and improving its technology. Each day, more people are getting connected to the online community.
The amount of time spent online by people is a key factor to consider regarding the future of e-commerce. The research indicates that people are now spending more time browsing online instead of spending time on other forms of entertainment. As a result of this trend, Internet shopping has been growing tremendously each year.

E-commerce opens the door to global consumers who can conduct business transactions 24 hours a day and 7 days a week from any location throughout the world. It is proven that people are likely to visit and shop online if an e-commerce website is culturally friendly. Cultural factors, therefore, are an important area on which to focus when designing an e-commerce website. Although based on limited numbers of journal articles, this paper provides a glimpse of the relationship between e-commerce, culture, and website design.

Therefore, it will be best practice to consider some of the following issues when designing an e-commerce website:

- Local cultural aspects to reflect the users' emotions, thoughts, desires, and most importantly the culture

A Narrative Literature Review and E-Commerce Website Research

- E-commerce website should be presented in the consumer's language to get better response and visitors
- Types of advertisement placed on the e-commerce website need to be taken in consideration
- Selecting appropriate culturally friendly colours when designing web page because colours used on the web page have diverse emotional and social consequences
- Understanding the local culture and customs when designing the e-commerce website

E-commerce research will definitely benefit both government and private sectors. The research on e-commerce will identify concerns regarding culture and e-commerce on which both the government sector and business organizations should focus to improve and invest. The research will also benefit academia; researchers, doctoral, and master's students will be able to use this study as reference material for further investigation in this area.

# References

Akhter, F. (2016). Cultural dimensions of behaviors towards e-commerce in a developing country context. *International Journal of Advanced Computer Science and Applications*, 7(4), Retrieved from https://thesai.org/Downloads/Volume7No4/Paper_13-Cultural_Dimensions_of_Behaviors_Towards_E_Commerce.pdf on February 26, 2018.

Asimionoaei, C. (2009). *Cultural issues in website design. A European perspective on electronic commerce.* Center for European Studies (CES) Working Papers, I, (1), Retrieved from http://ceswp.uaic.ro/articles/CESWP2009_I1_ASI.pdf on February 26, 2018.

Belkhamza, Z. and Wafa, S. (2014). The role of uncertainty avoidance on E-commerce acceptance across cultures. *International Business Research*, 7(5) Retrieved from http://ccsenet.org/journal/index.php/ibr/article/viewFile/32181/20469 on February 26, 2018.

Bhalekar, P., Ingle, S. and Pathak, K. (2014). The study of e-commerce. *Asian Journal of Computer Science And Information Technology,* 4(3) Retrieved from http://www.innovativejournal.in/index.php/ajcsit/article/viewFile/729/628 on February 26, 2018.

Brdesee, H., Corbitt, B. and Pittayachawan, S. (2012). Lessons to be learnt: Cultural means impeding e-commerce adoption in a Saudi industry. *International Journal of e-Education, e-Business, e-Management and e-Learning*, 2(6) Retrieved from http://www.ijeeee.org/Papers/169-G00003.pdf on February 26, 2018.

Dan, C. (2014). Electronic commerce: State-of-the-art. *American Journal of Intelligent Systems*, 4(4) Retrieved from http://article.sapub.org/10.5923.j.ajis.20140404.02.html#Sec5 on February 26, 2018.

Dehkordi, L.F., Shahnazari, A. and Noroozi, A. (2011). A study of the factors that influence the acceptance of e-commerce in developing countries: A comparative survey between Iran and United Arab Emirates. *Interdisciplinary Journal of Research in Business*, 1(6) Retrieved from https://pdfs.semanticscholar.org/c9ad/18a6d3ffc93352b3d643a1df0639fee52547.pdf on February 26, 2018.

Flood, M. (2014). *Humanism and culture*. Retrieved from http://humanistlife.org.uk/2014/06/16/humanism-and-culture/ February 26, 2018.

Gould, E.W. (2006). Results of a study on cultural values and e-commerce design for Malaysia and the United States. Retrieved from https://www.researchgate.net/publication/224059816_Results_of_a_Study_on_Cultural_Values_and_e-Commerce_Design_for_Malaysia_and_the_United_States on February 26, 2018.

Gupta, A. (2014). E-commerce: Role of e-commerce in today's business. *International Journal of Computing and Corporate Research* 4(1) Retrieved from https://www.ijccr.com/January2014/10.pdf on February 26, 2018.

Guzzo, R.A., Jackson, S.E. and Katzell, R.A. (1987). Meta-analysis analysis. *Research in Organizational Behavior* (9 Retrieved from https://smlr.rutgers.edu/sites/default/files/documents/faculty_staff_docs/MetaAnalysisAnalysis.pdf on February 26, 2018.

Hofstede, G. (1980). *Culture's consequences.* Beverly Hills, CA: Sage.

Hofstede, G. (1991). *Cultures and organizations: Software of the mind.* London: McGraw-Hill.

Kang, K.S. (2005). Considering the cultural issues of web design in implementing web-based e-commerce for international customers. *Proceedings of the Fifth International Conference on Electronic Business*, Hong Kong, 5-9 December 2005, pp. 323–327,

Khoshnampour, M. and Nosrati, M. (2011). An overview of E-commerce. *World Applied Programming*, 1(2) Retrieved from http://waprogramming.com/papers/50ae48c47a2ce6.65138277.pdf on February 26, 2018.

Khurana, A. and Mehra, J. (2015). E-commerce: Opportunities and challenges. *The International Journal Of Business & Management*, 3(1) Retrieved from http://www.theijbm.com/wp-content/uploads/2015/01/27.-BM1501-049.pdf on February 26, 2018.

King, W.R. and He, J. (2005). Understanding the role and methods of meta-analysis in IS research. *Communications of the Association for Information Systems*, 16(32) Retrieved from http://aisel.aisnet.org/cgi/viewcontent.cgi?article=3050&context=cais on February 26, 2018.

Li, M, Sun, X, Wang, H, Zhang, Y, Zhang, J (2011). Privacy-aware access control with trust management in web service. World Wide Web 14 (4), 407-430.

Lim, E. (2014). Adoption of E-Commerce in Manila. Paper presented at the DLSU Research Congress. 6-8 March 2014, Manila, Philippines. Retrieved from http://www.dlsu.edu.ph/conferences/dlsu_research_congress/2014/_pdf/proceedings/EBM-II-014-FT.pdf on February 26, 2018.

Lin, A.C.H. (2015). Facilitating cultural and creative industries to engage the Internet era: A new e-commerce strategic framework. *American Journal of Economics*, 5(5) Retrieved from http://article.sapub.org/10.5923.j.economics.20150505.14.html on February 26, 2018.

Lo, B.W.N. and Gong, P. (2005). Cultural impact on the design of e-commerce websites: Part I – site format and layout. *Issues in Information Systems*, 6(2) Retrieved from http://iacis.org/iis/2005/Lo_Gong.pdf on February 26, 2018.

Mohiuddin, M. (2014). Overview of the e-commerce in Bangladesh. *IOSR Journal of Business and Management*, 16(7) Retrieved from http://www.iosrjournals.org/iosr-jbm/papers/Vol16-issue7/Version-2/A016720106.pdf on February 26, 2018.

Nanehkaran, Y.A. (2013). An introduction to electronic commerce. *International Journal of Scientific & Technology Research*, 2(4) Retrieved from http://www.ijstr.org/final-print/apr2013/An-Introduction-To-Electronic-Commerce.pdf on February 26, 2018.

Nemat, R. (2011). Taking a look at different types of e-commerce. *World Applied Programming*, 1(2) Retrieved from http://www.waprogramming.com/index.php?action=journal&page=showpaper&jid=1&iid=2&pid=8 on February 26, 2018.

Ngai, E.W.T. and Wat, F.K.T. (2002). A literature review and classification of electronic commerce research. *Information & Management*, 39 Retrieved from https://pdfs.semanticscholar.org/7803/9359d134ccfc11724ef8af9da940462e54b8.pdf on February 26, 2018.

ORGANISATION FOR ECONOMIC CO-OPERATION AND DEVELOPMENT (OECD). (2002). *Measuring the information economy.* Retrieved from https://www.oecd.org/sti/ieconomy/1835738.pdf on February 26, 2018.

Pandey, D. and Agarwal, V. (2014). E-commerce transactions: An empirical study. *International Journal of Advanced Research in Computer Science and Software Engineering*, 4(3) Retrieved from https://www.researchgate.net/publication/291903546_E-commerce_Transactions_An_Empirical_Study on February 26, 2018.

Punhani, R. and Batra, S. (2014). Understanding cultural variations of e-commerce websites in a global framework. *Global Journal of Finance and Management*, 6(3) Retrieved from http://www.ripublication.com/gjfm-spl/gjfmv6n3_14.pdf on February 26, 2018.

Shahdadnejad, N. and Nakhaie, H. (2011). The role of culture in the development of electronic commerce. *3rd International Conference on Information and Financial Engineering,* Singapore, pp. 513-517.

Shahriari, S., Shahriari, M. and Gheiji, S. (2015). E-commerce and it impacts on global trend and market. *International Journal of Research – Granthaalayah*, 3(4) Retrieved from http://granthaalayah.com/Articles/Vol3Iss4/03_IJRG15_A04_63.pdf on February 26, 2018.

Singh, N., Alhorr, H.S. and Bartikowski, B.P. (2010). Global e-commerce: A portal bridging the world markets. *Journal of Electronic Commerce Research*, 11(1), Retrieved from http://web.csulb.edu/journals/jecr/issues/20101/paper0.pdf on February 26, 2018.

Wang, H, Cao, J, and Zhang, Y (2001). A consumer scalable anonymity payment scheme with role based access control. Web Information Systems Engineering, 2001, Proceedings of the Second International Conference, 3-6 Dec. 2001, Kyoto, Japan.

Wang, H, Cao, J, and Zhang, Y (2005). A flexible payment scheme and its role-based access Control. IEEE Transactions on knowledge and Data Engineering 17 (3), pp. 425-436.

Wang, H, Cao, J, and Zhang, Y (2002). Ticket-based service access scheme for mobile users. Australian Computer Science Communications 24 (1), Volume 24 Issue 1, January-February, 2002, Page 285-292.

Wang, H and Zhang, Y (2011). Untraceable off-line electronic cash flow in e-commerce. Australian Computer Science Communications 23 (1), pp. 191-198.

Wang, H, Zhang, Y, Cao, J (2009). Effective collaboration with information sharing in virtual universities. IEEE Transactions on Knowledge and Data Engineering 21 (6), pp. 840-853.

Webster, J. and Watson, R.T. (2002). Analyzing the past to prepare for the future: Writing a literature review. *MIS Quarterly* 26(2) Retrieved from http://www.magnusvg.com/study/wp-content/uploads/2011/02/Analyzing-the-past-to-prepara-for-the-future.pdf on February 26, 2018.

Yongrui, Q., Sheng, Z., Falkner, N. J. G., Dustdar, S., Wang, H., and Vasilakos, A. V. (2014). When Things Matter: A Data-Centric View of the Internet of Things. CoRR, 2014, volume abs/1407.2704, Retrieved from https://arxiv.org/pdf/1407.2704.pdf on February 26, 2018.

Zwass, V. (1996). Electronic commerce: Structures and issues. *International Journal of Electronic Commerce*, 1(1) Retrieved from https://pdfs.semanticscholar.org/9f36/c5cea1838378b66b102c6afb996b77e78233.pdf on February 26, 2018.